\documentclass{article}
\usepackage{authblk}
\usepackage[english]{babel}

\usepackage[letterpaper,top=2cm,bottom=2cm,left=3cm,right=3cm,marginparwidth=1.75cm]{geometry}

\usepackage{amsmath}
\usepackage{graphicx}
\usepackage[colorlinks=true, allcolors=blue]{hyperref}

\usepackage[utf8]{inputenc}

\title{Optical Channel Aggregation by Coherent Spectral Superposition with Electro-Optic Modulators}
\author[1]{Arijit~Misra}
\author[1]{Stefan~Preu{\ss}ler}
\author[1]{Karanveer~Singh}
\author[1]{Janosch~Meier}
\author[1]{Thomas~Schneider}
\affil[1]{THz Photonics Group, Technische Universit\"at Braunschweig, Schleinitzstra{\ss}e 22, 38106 Braunschweig, Germany. (arijit.misra@ihf.tu-bs.de)}
\date{} 
\begin{document}
\maketitle

\begin{abstract}
As the bit rates of routed data streams exceed the throughput of single wavelength-division multiplexing channels, spectral traffic aggregation becomes essential for optical network scaling. Here we propose a scheme for all-optical aggregation of several low bitrate channels to fewer channels with higher spectral efficiency. The method is based on optical vector summation facilitated by coherent spectral superposition. Thereby it does not need any optical nonlinearities and is based on linear signal processing with an electro-optic modulator. Furthermore, optical phase tuning required for vector addition can be easily achieved by a phase tuning of the radio frequency signal driving the modulator. We experimentally demonstrate the aggregation of two 10 Gbaud BPSK signals into one 10 Gbaud QPSK and one 10 Gbaud PAM-4 signal, the aggregation of two 10 Gbaud QPSK signals into 10 Gbaud QAM-16, as well as the aggregation of sinc-shaped Nyquist signals. The presented concept of in-line all-optical aggregation demonstrates considerable improvement in network spectrum utilization and can significantly enhance the operational capacity with reduced complexity. It provides a new way for realizing the flexible optical transmission of advanced modulation format signals, and suits for future dynamically reconfigurable optical networks. Since the method is based on linear signal processing with electro-optic modulator, integration into any integrated photonic platform is straight-forward.
\end{abstract}

\section{Introduction}

\label{sec:intro} 
To satisfy the sustained growth in demand for communication bandwidth, the use of advanced modulation formats like QAM has attracted abundant interest for high spectral efficiency transmission \cite{Winzer2012}. The generation of an M-QAM optical signal with a symbol rate of $R$ GBd requires two digital to analog converters (DACs), each with a resolution and sampling rate of at least $\log_2 \sqrt[]{M}$ and $R$ GS/s respectively. These electronic DACs at high baud rates have finite linearity ranges, which leads to inconsistencies in the data constellation points. Additionally, they increase the complexity and power consumption of the network. Instead,  all-optical signal processing techniques enable high data rate operation with flexible and efficient bandwidth utilization with low power consumption\cite{Willner2019}. Specifically, the aggregation of two or more less complex, lower bit rate channels to a single higher bit rate channel can be performed all optically through coherent vector addition. Channel aggregation enables promising advances for the elastic optical networking (EON) \cite{Liu2019a,Liu2019b} and near future fifth-generation (5G) technology \cite{Musumeci2016} with the existing network architecture. Thus, aggregation gives an easy option to switch between simple and easy to generate modulation formats including OOK and BPSK in metro and local access networks (M/LAN) and higher modulation formats like QAM with improved transmission capacity in optical backbone networks.\par
Commonly, different nonlinear optical effects like four wave mixing (FWM), cross phase modulation (XPM), self-phase modulation (SPM), parametric amplification (PA), and cross gain modulation (XGM) arising from the second or third order susceptibility of highly nonlinear fiber (HNLF), periodically poled lithium niobate (PPLN) waveguide or semiconductor optical amplifiers (SOAs) have been used for channel aggregation \cite{Liu2019a,Liu2019b,Musumeci2016,Qiankun2021,Lu2015,Huang2011,Zhang2013,Chitgarha2013,Chitgarha2014,Fallahpour2020,Amano2019}. In \cite{Liu2019a,Liu2019b,Amano2019} for instance, aggregation of OOK and MPSK signals to a higher bit rate QPSK signals based on XPM using HNLF have been shown. QAM-8 signals have been generated by the aggregation of QPSK and OOK signals using XPM and XGM in SOA \cite{Amano2019} and also from BPSK inputs using XPM and PA in HNLF \cite{Qiankun2021,Lu2015}. Optical aggregation of four OOK to a single QAM-16 signal using a nonlinear optical loop mirror based on XPM and parametric amplification has been presented in \cite{Huang2011}. To demonstrate tunable optical aggregation of two QPSK channels to a QAM-16 channel based on nonlinear wave mixing, many different coherent comb sources like mode-locked lasers (MLL) \cite{Chitgarha2014} and Kerr combs originating from microring resonators \cite{Fallahpour2020} have been used. These methods also use nonlinear wave mixing for the coherent vector summation of the input channels. In a recent experiment, aggregation of sinc-shaped Nyquist channels has been reported using 2\textsuperscript{nd} order nonlinearity of PPLN waveguides \cite{Fallahpour2020}. \par
All channel aggregation methods shown so far require various nonlinear effects, special waveguides, pump laser sources, and sufficient optical power for the nonlinear interactions, leading to increased system complexity, polarization dependence, power consumption, and additional noise, which hinders the operational reconfigurability and range of applications. It would be of great interest to have a less complex linear approach for channel aggregation which can be directly deployed in elastic optical networks and offers a potential to implement on an integrated photonic platform like silicon photonics. \par
In this paper, we demonstrate a simple technique for flexible all-optical aggregation without using any optical nonlinearity or specific waveguide. It uses a completely linear approach for the vector summation of the input channels with low modulation formats into fewer channels with higher modulation formats. The signal channels to be aggregated must be modulated on coherent carriers at distinct wavelengths, which can be achieved by Kerr combs using ring resonators, MLLs, or electro-optic modulators \cite{Chitgarha2014,Fallahpour2020}. Then a phase or intensity modulator modulates the input channels to generate sidebands in a way that they superimpose for phase coherent vector summation. Via a control over the phase of the radio frequency (RF) input to the modulator, the phase difference between the superimposed sidebands can be controlled. We demonstrate experimental results for 10 Gbaud QAM-16 signal generation from aggregation of two 10 Gbaud QPSK signals. Two 10 Gbaud BPSK signals are aggregated to one 10 Gbaud QPSK and also one 10 Gbaud PAM-4 signal. The aggregation of sinc-shaped Nyquist channels of lower modulation formats to higher moduation format Nyquist channel is also presented experimentally.\par
This new concept of in line all-optical aggregation can significantly enhance the operational capacity with considerably reduced complexity. Moreover, the utilization of standard components like electro-optic modulators offered as a part of the process design kits of commercial foundries, provides an opportunity of on-chip implementation with low cost and high yield.\par
This article is structured as follows. In Section II, we explain the operating principle. In Section III, we describe the experimental setup and present the results. Finally, after a brief discussion about the concept and experimental results in Section IV, we conclude this paper in Section V. 
%%%%%%%%%%%%%%%%%%%%%%%%%%%%%%%%%%%%%%%%%%%%%%%%%%%%%%%%%%%%

\section{OPERATING PRINCIPLE}
\label{sec:principles}
\begin{figure}[!htb]
   \centering
   \includegraphics[width=0.9\columnwidth]{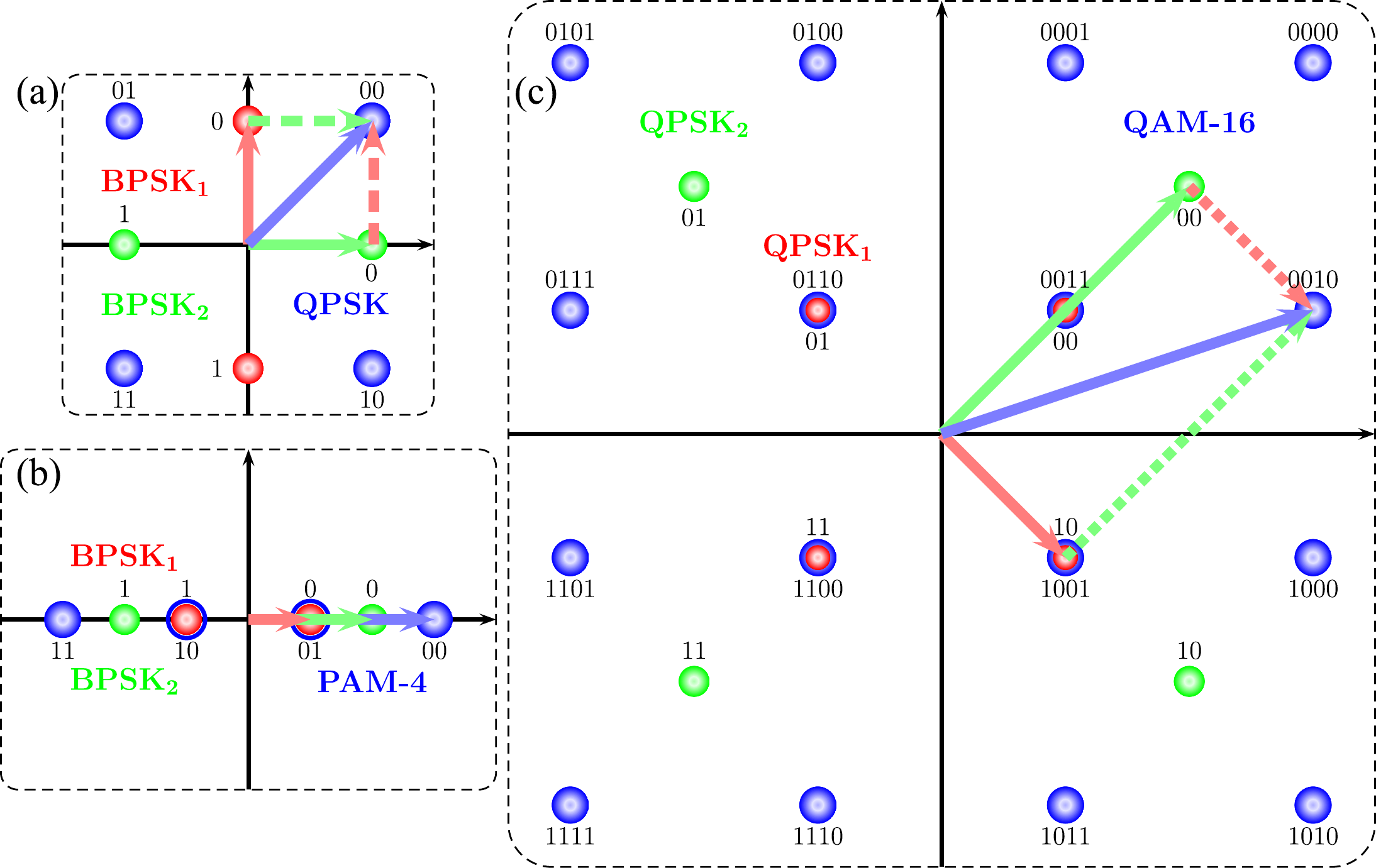}
   \caption{Schematic representation of the aggregation of two BPSK signals as shown with the red and green constellations to (a) one QPSK, and (b) one PAM-4 signal as shown in blue. For QPSK, both the parent BPSK signals have same peak to peak amplitude ($\alpha=1.0$, with $\alpha$ as the amplitude ratio) but the corresponding carriers are $90^\circ$ out of phase. For PAM-4 the carriers are in phase but, BPSK\textsubscript{1} has half the amplitude of BPSK\textsubscript{2} ($\alpha=0.5$). (c) Two QPSK signals with QPSK\textsubscript{1} (red) and QPSK\textsubscript{2} with $\alpha=0.5$ can be aggregated to one QAM-16 signal (blue) through vector summation. The arrows show the vector summation  of the 00 symbol from QPSK\textsubscript{2} and the 10 symbol from QPSK\textsubscript{1} to form the 0010 symbol of the QAM-16 signal.  Other modulation formats like M-PSK, M-ASK, and M-QAM with higher values of M, hence higher spectral efficiency can be achieved in the same way.}
   \label{fig:vector} 
\end{figure}

The all-optical aggregation of lower bit rate channels is based on optical vector summation as shown in Fig. \ref{fig:vector}. In Fig. \ref{fig:vector}(a) the vector for the symbol 0 from BPSK\textsubscript{1} (red) is summed up with the vector for the 0 symbol from BPSK\textsubscript{2} (green) to build the 00 vector of the QPSK signal (blue). To achieve this the two vectors should have the same amplitude but a phase difference of $\phi=\pi/2$. If the two vectors have the same phase but a different amplitude, a PAM-4 signal can be generated, as illustrated in Fig. \ref{fig:vector}(b). Similarly, Fig. \ref{fig:vector}(c) shows the generation of a QAM-16 signal by vector summation of two QPSK signals. \par

\begin{figure*}[!htb]
   \centering
   \includegraphics[width=0.75\textwidth]{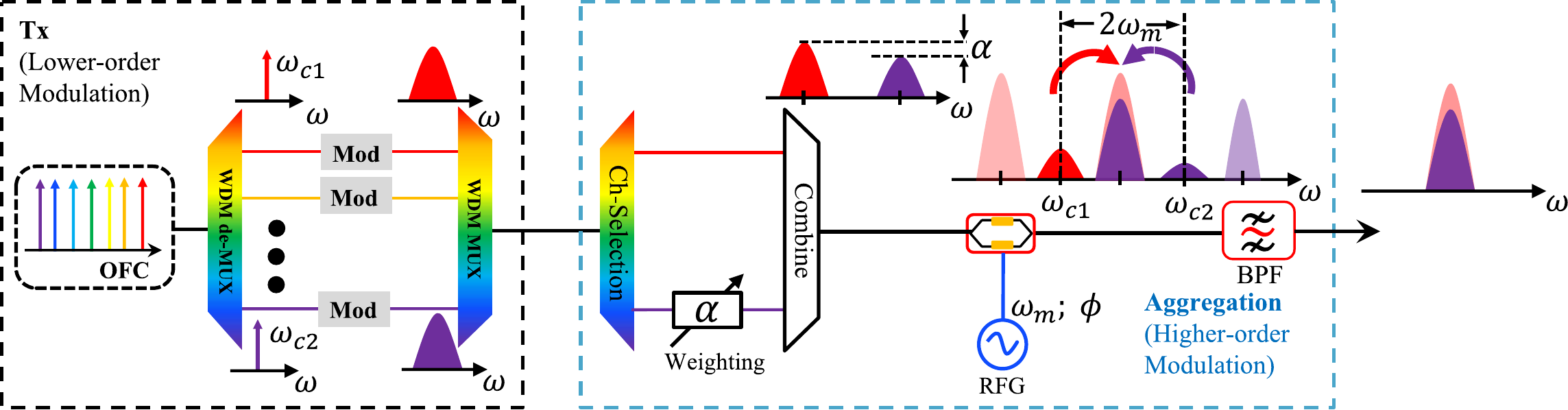}
   \caption{Graphical illustration of the operating principle of the all optical aggregation method using electro-optic modulators. In principle several input channels with lower spectral efficiency can be aggregated to fewer output channels with a higher spectral efficiency. Here the method is illustrated for two independent lower bit rate optical data channels around $\omega_{c1}$ and $\omega_{c2}$, generated by modulation of a phase coherent optical frequency comb. These two channels are aggregated to build an output channel with twice the spectral efficiency by superposition with a correct phase and amplitude difference between them. A weight factor ($\alpha$) assigns different amplitudes to the vectors to be summed. The lower sideband (LSB) from the higher frequency channel ($\omega_{c2}$) overlaps with the higher sideband (HSB) from the lower frequency channel ($\omega_{c1}$) after modulation by the electro-optic modulator. The phase difference required for the vector summation is achieved by varying the input RF phase $\phi$ to the modulator generated from a RF generator (RFG). Here an MZM is employed in carrier suppression ($V_{\mathrm{bias}}=V_\pi$) for better illustration. A bandpass filter (BPF) then allows the required superimposed band to be transmitted after aggregation.}
   \label{fig:concept} 
\end{figure*}

Optical aggregation using coherent vector addition \cite{Chitgarha2013,Chitgarha2014,Fallahpour2020} can be achieved by coherently superposing two signal spectra \cite{Parmigiani2012} with suitable amplitude and phase difference in the carriers. The operating principle of coherent spectral superposition with optical modulators, proposed here, is illustrated in Fig. \ref{fig:concept}. The coherent superposition does not rely on nonlinear optical effects. In contrast, electro-optic modulation is utilized. It essentially relies on an optical frequency comb with coherent comb lines at the transmitter. For the method the comb lines do not need to have the same or a linear phase depency to each other, it is sufficient to have a stable phase relationship. Therefore, a wide variety of comb sources like electro-optic modulators, MLLs, or kerr comb generators like microresonators can be utilized. Such frequency comb sources are increasingly becoming an integral part of high-capacity coherent optical communication systems \cite{Hu2018,Corcoran2020}. 
\par

In principle, several channels with lower spectral efficiencies can be combined to construct fewer channels with higher spectral efficiency. However, in Fig. \ref{fig:concept} the aggregation of two channels to one output channel is shown for the sake of simplicity. Carriers originating from a phase locked comb are modulated with low spectral efficiency data signals in a less complex WDM transmitter as shown in the left block.  Afterwards, in the right block, which might be located in a network node, any two channels are aggregated to build a channel with twice the spectral efficiency, preserving the spectral width. At first, the two targeted input channels at $\omega_{c1}$ and $\omega_{c2}$ are chosen by an WDM filter. One of the channels is then weighted in amplitude with a suitable weight factor $\alpha$ determined by the targeted higher modulation format. These two channels are then combined and injected into an electro-optic modulator.  The modulator is driven with a sinusoidal RF signal of frequency $\omega_m$. Due to modulation, the lower sideband (LSB) resulting from the channel around $\omega_{c2}$ is superimposed with the higher sideband (HSB) resulting from the channel around $\omega_{c1}$. As indicated in Fig. \ref{fig:concept} the frequency spacing between the parent channels can be as high as $2\omega_m$. To achieve the required vector summation for the intended modulation format, merely a superposition of the spectral amplitudes is insufficient. The optical carrier phases of the superimposing spectra are also an important factor to be considered. For the presented concept the optical carrier phases of the superposing spectra are manipulated by the phase of the input sinusoidal RF signal to the modulator. Following, we have presented a mathematical analysis regarding the phase relationship of the two optical sidebands resulting from the electro-optic modulation process which in turn determines the feasibility of the proposed concept.\par

In general, a complex quadrature amplitude modulated carrier can be decomposed into two amplitude modulated signals known as in-phase (I) and quadrature (Q) components. These two components are  related by a constant phase offset of $\pi/2$ between the respective carrier wave ($\omega_c$).\par
Let us consider the in-phase component can be expressed as, 
\begin{equation}
\centering
s_I(t)=I(t)\cdot \cos{\Big(\omega_c\left(t+t_a\right)\Big)}.
\label{Eq:1}
\end{equation}
Here, $t_a$ is a time constant defining the phase of the I-component. After being modulated by a single drive MZM with a sinusoidal RF signal with angular frequency $\omega_m$, the output can be written as,
\begin{equation}
\begin{aligned}
&s_I^M(t) \\
&=\frac{1}{\sqrt[]{2}} I(t)\cdot \mathrm{cos}\Big(\omega_c(t+ A\cdot \mathrm{sin}(\omega_m(t+t_p))+b+t_a)\Big)\\
&+\frac{1}{\sqrt[]{2}}I(t)\cdot \cos{\Big(\omega_c(t+t_a)\Big)}
\end{aligned}
\label{Eq:2}
\end{equation}
with $t_p$ as an arbitrary time shift of the RF input with amplitude $A$, and $b$ is the time shift induced by the dc bias to the modulator. The time shift $t_p$ can be realized simply by an electrical phase shift of the RF input.
Using the Jacobi-Anger expansion and expressing the associated constant phase term as $\theta=\omega_c b+ \omega_c t_a$,
\begin{equation}
\begin{aligned}
s_I^M(t)&=\frac{1}{\sqrt[]{2}} I(t)\cdot \mathrm{Re}\Big(e^{i(\omega_c t+ \theta)} \\
&\hspace{1in} \cdot \sum_{n=-\infty}^{\infty} J_n(A\cdot \omega_c) \cdot e^{i\cdot n(\omega_m t+\omega_m t_p)}\Big)\\
&+\frac{1}{\sqrt[]{2}}I(t)\cdot \cos{(\omega_c(t+t_a))}.
\end{aligned}
\label{Eq:3}
\end{equation}
Here, the summation variable $n$ corresponds to the order of the sidebands. Considering only the carrier and 1\textsuperscript{st} order sidebands ($n=0,\pm 1$),
\begin{equation}
\begin{aligned}
&\tilde{s}_I^M(t)\\
&=\frac{1}{\sqrt[]{2}} I(t) \cdot \Big[ J_0(A\cdot \omega_c) \cdot \mathrm{cos}(\omega_ct+\theta)\\
&\hspace{0.2in} +J_{+1}(A\cdot \omega_c)\cdot\mathrm{cos}\big((\omega_c+\omega_m)t+\omega_m t_p +\theta\big)\\
&\hspace{0.2in}+ J_{-1}(A\cdot \omega_c)\cdot\mathrm{cos}\big((\omega_c-\omega_m)t-\omega_m t_p +\theta\big)\Big]\\
&+\frac{1}{\sqrt[]{2}}I(t)\cdot \cos{(\omega_c(t+t_a))}.
\end{aligned}
\label{Eq:4}
\end{equation}
It can be seen from Eq. \ref{Eq:4} that the LSB ($n=-1$) and HSB ($n=+1$) exhibit phase change the inverse direction, when the phase of the input RF signal to the modulator ($\phi=\omega_m t_p$)  is varied. Therefore, with a careful adjustment of the RF phase an arbitrary phase relationship between the two optical sidebands (LSB and HSB) can be achieved, resulting in an adjustable optical phase relationship between the the two input data channels. \par
A superposition of the HSB and LSB originating from $\omega_{c1}$ and $\omega_{c2}$ respectively, will lead to the aggregated I component as:
\begin{equation}
\begin{aligned}
&s_I^A(t)\\
&=\frac{1}{\sqrt[]{2}} \Big[I_1(t) \cdot J_{+1}(A\cdot \omega_{c1})\cdot\mathrm{cos}\big((\omega_{c1}+\omega_m)t+\omega_m t_p +\theta_1\big)\\
&\hspace{0.2in}+I_2(t) \cdot J_{-1}(A\cdot \omega_{c2})\cdot\mathrm{cos}\big((\omega_{c2}-\omega_m)t-\omega_m t_p +\theta_2\big)\Big]
\end{aligned}
\label{Eq:5}
\end{equation}
Here we have expressed the constant phase terms associated with the I-components as, $\theta_1=\omega_c b+ \omega_c t_{a1}$ and $\theta_2=\omega_c b+ \omega_c t_{a2}$ respectively.
Equation \ref{Eq:5} indicates that the phase ($\omega_c t_a$) associated with the I-component is still preserved after the electro-optic modulation along with a constant phase contribution from the bias of the modulator ($\omega_c b$). Hence, it can be safely assumed that the above treatment will be valid for the quadrature component (Q) as well. For the Q-component the phase change due to the modulation will also be exactly the same amount as for the I-component. However, the constant carrier phase difference of $\pi/2$ between the I and Q components will remain preserved. \par

\begin{figure*}[!htb]
   \centering
   \includegraphics[width=0.75\textwidth]{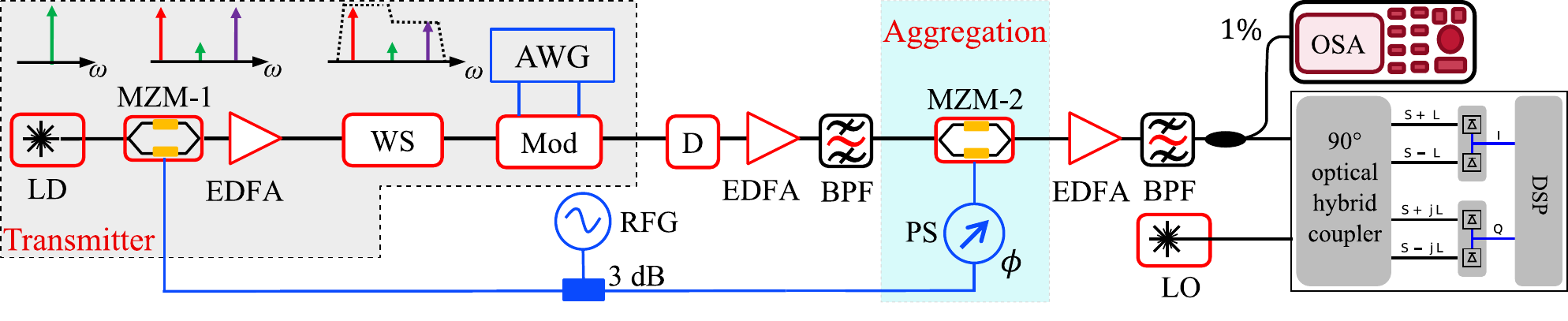}
   \caption{The proof-of-concept experimental set up for the all-optical aggregation. An optical carrier wave generated from a laser diode is modulated by MZM-1 in the carrier suppression mode to generate two phase locked ancillary carriers. A waveshaper (WS) introduces different amplitudes to the carriers, defined by $\alpha$. Next an I-Q modulator (Mod) modulates the carriers with the same data signal generated by an arbitrary waveform generator (AWG). To decorrelate the data content, a dispersion module (D) introduces a delay between the two channels. The vector summation of the two channels is accomplished by MZM-2. A radio frequency generator (RFG) generates the RF frequency inputs to both MZMs and an electrical phase shifter (PS) is used for the phase manipulation. A heterodyne optical coherent receiver with external local oscillator (LO) has been used to receive the aggregated signal. Erbium-doped fiber amplifiers (EDFAs), together with bandpass filters (BPFs) to suppress the out of band spontaneous emission noise, have been used to amplify the optical signals if needed.}
   \label{fig:setup} 
\end{figure*}

As shown in Fig. \ref{fig:concept} the two superimposing spectra originate from the modulation of the two different parent channels. One of them is the LSB while the other is the HSB generated by the same input RF signal applied to the modulator. Following from the above equations, the optical phases of the corresponding  carriers change by equal amount in opposite directions ($\pm\omega_m t_p$). Therefore, by controlling the RF phase the phase difference between the corresponding optical carriers associated with the superimposing spectra can be arbitrarily regulated to get the required vector summation for the intended modulation format.

%%%%%%%%%%%%%%%%%%%%%%%%%%%%%%%%%%%%%%%%%%%%%%%%

\section{Experiment and Results}
\label{sec:exp}
To verify the feasibility of the proposed aggregation method, proof-of-concept experiments were carried out. In this section the experimental details are discussed first, then the results are presented.
 \subsection{Experiment}
 The experimental arrangement is illustrated in Fig. \ref{fig:setup}. In absence of a comb source, two ancillary carriers were generated from one primary carrier via electro-optic modulation. The output from a laser diode (LD) emitting at 193.4 THz was modulated by MZM-1 (Optilab-IM1550-20a) in carrier suppression with an RF input of 18 GHz to generate two sidebands 36 GHz apart. The primary carrier at 193.4 THz was suppressed by 20 dB by adjusting the MZM-1 bias to the minimum transmission point. A wave shaper (WS, Finisar-1000s) was used as a programmable filter to have different weights for the vector addition. Only the power in the two ancillary carriers were independently adjusted by the WS without any phase adjustment. \par 
An I-Q modulator modulated the carriers with desired lower spectral efficiency modulation formats. Same data was modulated over both the channels. Therefore, to decorrelate them, relative symbol delays were applied between the two channels using a Teraxion dispersion-compensating fiber module (D). The dispersion induced by the module can be tuned with a resolution of $\pm10$ ps/nm. The dispersion module changes the relative phase between the two channels as well. However, since the relative phase for the aggregation is adjusted with the RF signal driving MZM-2, this had no effect on the experiment.\par
\begin{figure}[!hbt]
   \centering
   \includegraphics[width=0.75\columnwidth]{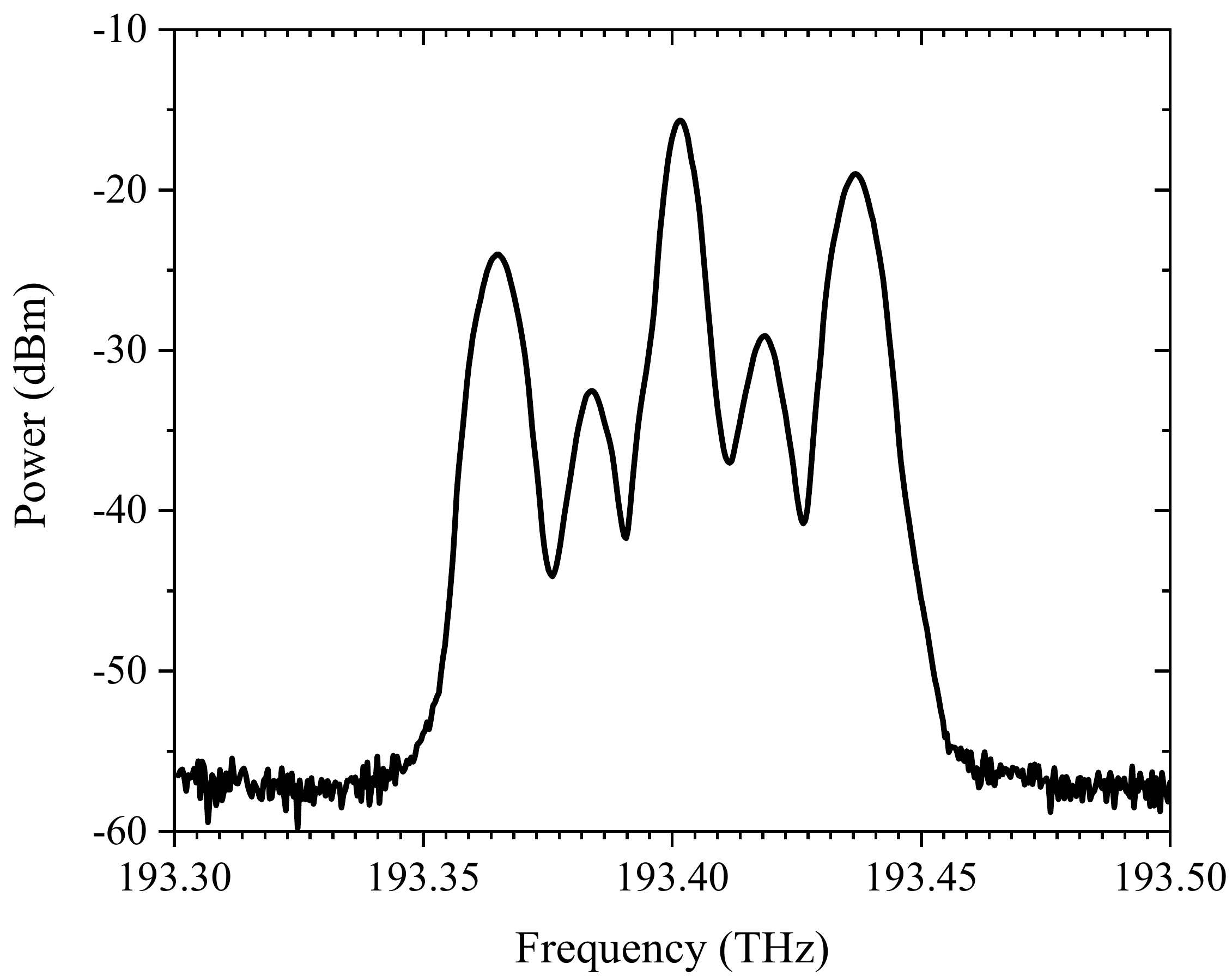}
   \caption{Experimental optical spectrum of the output signal after MZM-2. The ancillary channels were modulated with 10 GBd QPSK signal. Hence, the aggregated signal was of 10 GBd QAM-16 modulation format.}
   \label{fig:spec} 
\end{figure}

The aggregation of the two channels, or their coherent superposition were carried out by MZM-2, which is driven with the same 18 GHz RF as MZM-1. Therefore, as depicted in Fig. \ref{fig:concept}, the upper sideband from the lower carrier is superimposed with the lower sideband from the carrier with the higher frequency at the frequency position of the laser output at 193.4 THz. The spectrum as measured by an optical spectrum analyzer (OSA) as shown in Fig. \ref{fig:spec}. As the two superposing spectra are from two different sidebands we have a relative phase relationship between them. A tunable RF phase shifter (PS) is used to achieve this phase tuning between the two overlapping spectra. Since the RF phase alters the phase of the optical waves in opposite directions, as shown in Sec. \ref{sec:principles}, any phase relationship between the optical channels can be adjusted to generate the required vector sum. EDFAs along with a 3 nm bandpass filter (BPF) were used to amplify the signals and suppress the out of band amplified spontaneous emission noise.\par
For the demodulation of the aggregated optical channel an optical coherent detector was used with another laser source as the local oscillator (LO). For the sake of experimental simplicity, the required bandpass filtering of the aggregated optical channel was alternatively done in the electrical domain via lowpass filtering after demodulation. A coherent modulation analyzer (Tektronix-OM1106) performed the required digital signal processing (DSP) of the recorded waveforms in a real time oscilloscope (Tektronix DPO73304) for the visualization of symbol constellations and the measurement of other performance metrics like Q-factor and error vector magnitude (EVM).

\subsection{Results} 
The experiments were carried out for two types of signal formats. The first one being standard lower order non-return to zero (NRZ) modulation format signals like BPSK or QPSK with raised cosine spectral pre-shaping. The full width at half maximum (FWHM) of the spectrum corresponds to the symbol rate with a roll-off factor of 1.0. The second one being Nyquist BPSK or QPSK, where symbols were modulated on sinc-shaped Nyquist pulse sequences \cite{Soto2013,SotoCleo,Misra2021}. Thus, the spectrum of each channel is confined in a rectangular bandwidth (zero roll-off factor), enabling the transmission with the maximum possible symbol rate \cite{Misra2021}.

\subsubsection{Aggregation of standard signal formats}
\begin{figure}[!hbt]
   \centering
   \includegraphics[width=\columnwidth]{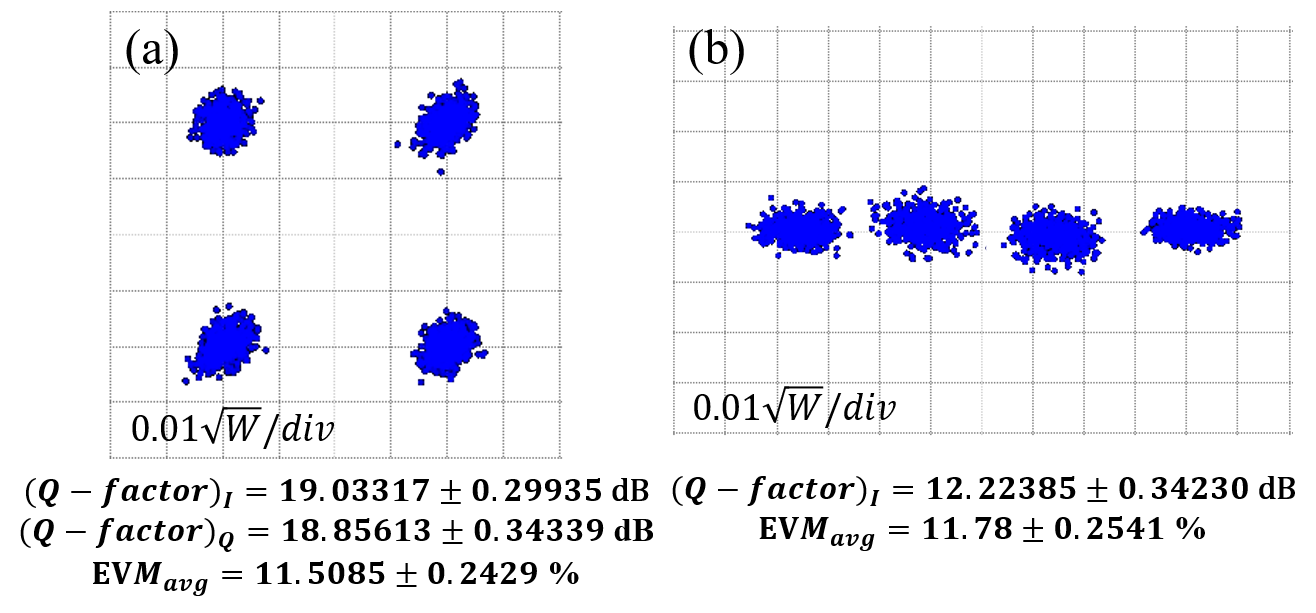}
   \caption{Symbol constellations arising from the aggregation of two 10 GBd BPSK signals to 10 GBd QPSK (a) and 10 GBd PAM-4 (b) signals. The measured Q-factors and average EVMs are presented as well. A back to back measurement of one of the parent BPSK channels resulted in, \textit{Q-factor}$=19.9933\pm0.24689$ and \textit{EVM\textsubscript{avg}}$=12.7211\pm0.5277$.}
   \label{fig:normal1} 
\end{figure}

In the first experiments, the aggregation of two 10 GBd BPSK channels to one 10 GBd PAM-4 or one 10 GBd QPSK channel and two 10 GBd QPSK channels to one 10 GBd QAM-16 has been demonstrated experimentally. The DCF module induced a dispersion of -270 ps/nm for delaying the two parent channels with respect to each other. Fig. \ref{fig:normal1} depicts the symbol constellation diagrams for the output 10 GBd QPSK (in Fig. \ref{fig:normal1}(a)) and 10 GBd PAM-4 (in Fig. \ref{fig:normal1}(b)) signals. For Fig. \ref{fig:normal1}(a) the amplitude of both input channels has to be equal ($\alpha=1$) with the carrier phase difference of $\phi=90^\circ$. Correspondingly, for a PAM-4 modulation as in Fig. \ref{fig:normal1}(b) $\alpha=0.5$ and $\phi=0^\circ$. The measured average error vector magnitudes and Q-factors are listed as well.\par
\begin{figure}[!hbt]
   \centering
   \includegraphics[width=\columnwidth]{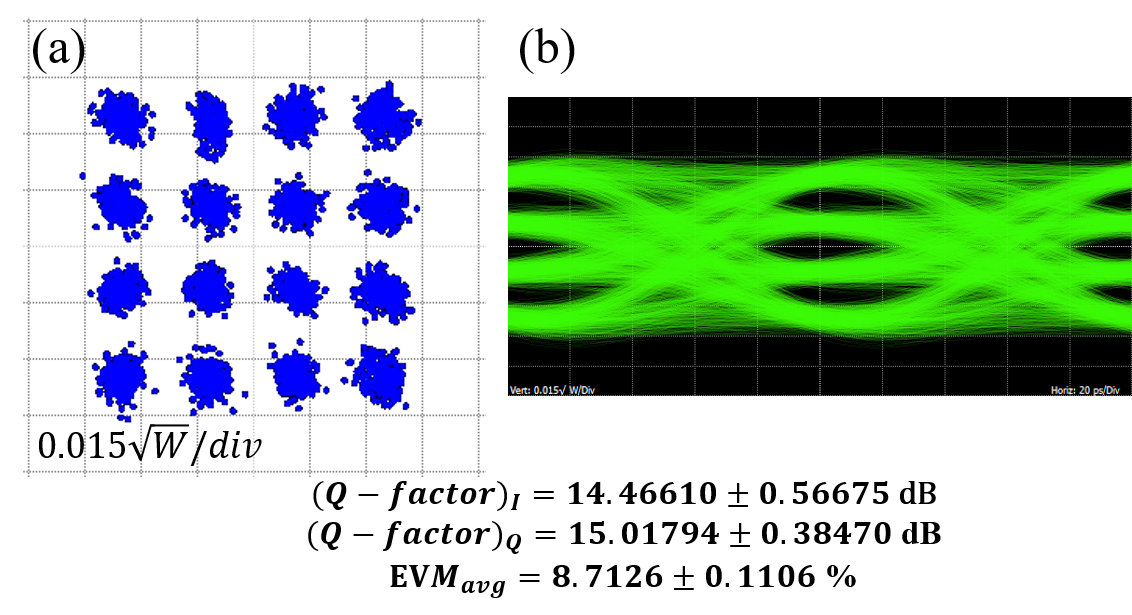}
   \caption{(a) Generated 10 GBd QAM-16 symbol constellation from the aggregation of two 10 GBd QPSK channels. A back to back measurement of one of the parent QPSK channels resulted in, \textit{Q-factor}$=20.366\pm0.19685$ and \textit{EVM\textsubscript{avg}}$=12.88\pm0.1646$. (b) The measured eye diagram of the corresponding in-phase component. Measured performance metrics are presented.}
   \label{fig:normal2} 
\end{figure}

The result for two 10 GBd QPSK channels aggregated to form a QAM-16 signal is depicted in Fig. \ref{fig:normal2}(a). the measured eye diagram of the in-phase (I) component of the aggregated signal has been shown in Fig. \ref{fig:normal2}(b). Fig. \ref{fig:spec} shows the coressponding optical spectrum after MZM-2.
\subsubsection{Aggregation of Nyquist channels}
Since they have rectangular spectra, sinc-shaped Nyquist signals offer the possibility to transmit data with the maximum possible symbol rate in a given bandwidth and they enable wavelength division multiplexing without any guard-band between the channels \cite{Nakazawa2012}. However, sinc-shaped Nyquist pulses are unlimited in the time domain and thus just a mathematical construct. Thus, usually complex digital \cite{Schmogrow12} or optical \cite{Nakazawa2014} signal processing is required for the generation and detection of the signals. Recently a new method for Nyquist WDM based on sinc pulse sequences generated, modulated and multiplexed by a single modulator and demultiplexed by another single modulator, which drastically reduces the requirements for digital signal processing has been proposed \cite{Soto2013,Misra2021}.  Here we have followed the same method to generate signals to be aggregated which are modulated on sinc-sequences with two zero crossings. The experiments were performed using 5 and 8 GBd BPSK and QPSK signals as parent channels. The induced dispersion required for the decorrelation of 5 and 8 GBd Nyquist signals were  -520 ps/nm and -440 ps/nm respectively.\par
\begin{figure}[!hbt]
   \centering
   \includegraphics[width=\columnwidth]{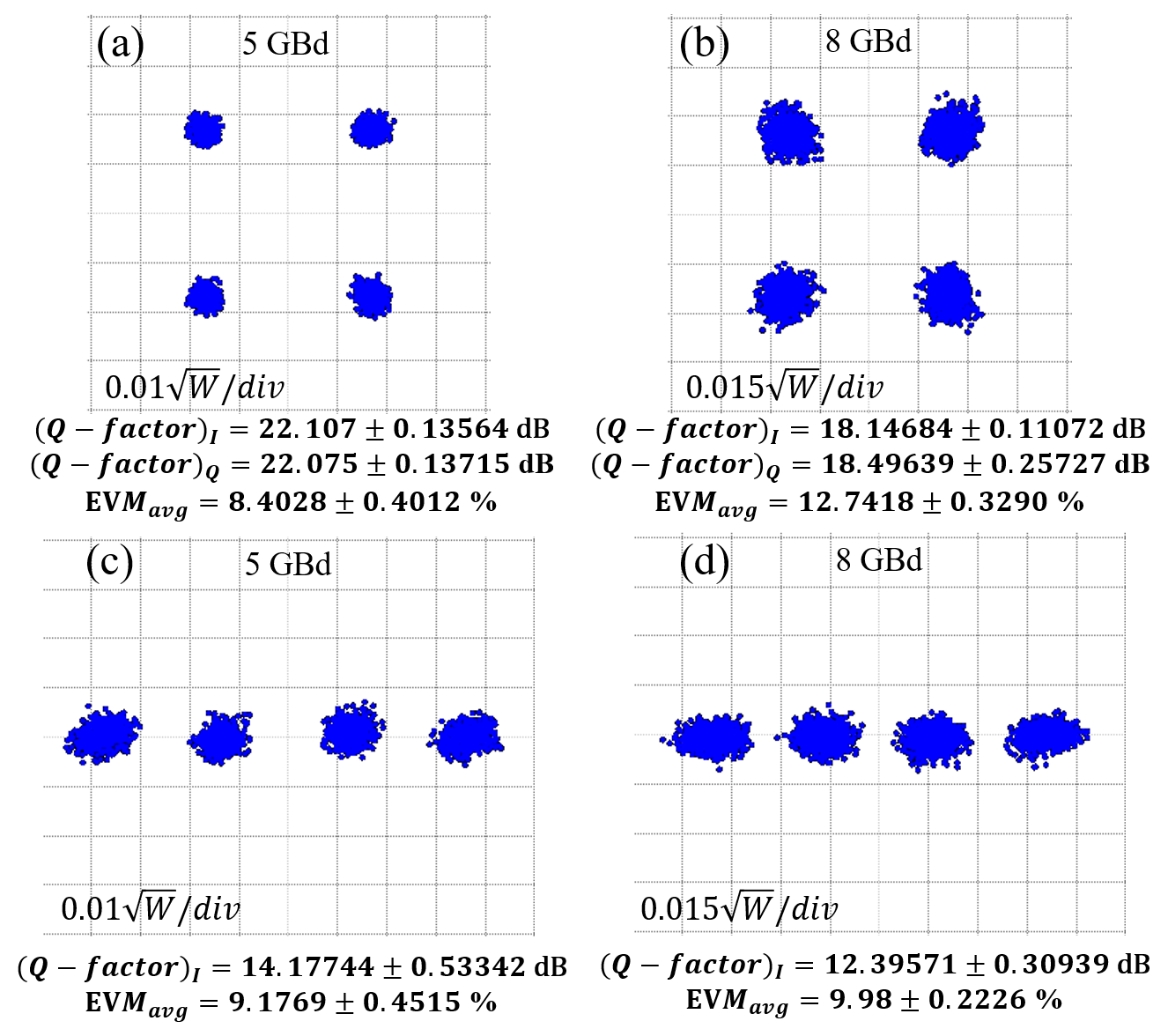}
   \caption{Symbol constellation of the generated (a) 5 GBd Nyquist QPSK, (b) 8 GBd Nyquist QPSK, (c) 5 GBd Nyquist PAM-4, and (d) 8 GBd Nyquist PAM-4 signals aggregated from two Nyquist BPSK signals of corresponding symbol rates. The \textit{Q-factor} and \textit{EVM\textsubscript{avg}} of one of the parent Nyquist BPSK channels at 8 GBd were found to be $22.1279\pm0.3654$ and $9.43\pm0.1493$ respectively for a back to back measurement.}
   \label{fig:sinc1} 
\end{figure}
\begin{figure}[!hbt]
   \centering
   \includegraphics[width=\columnwidth]{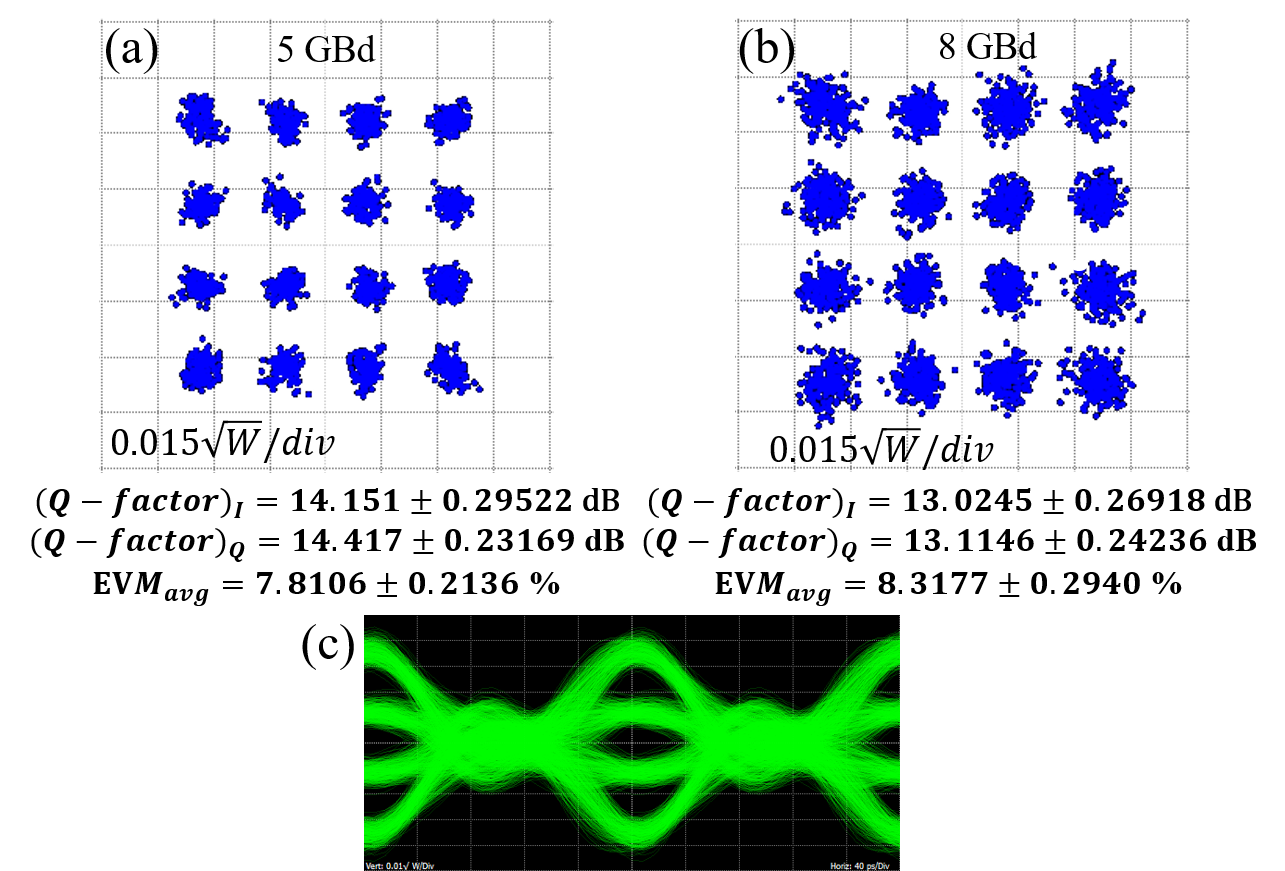}
   \caption{Symbol constellation and performance metrics of generated (a) 5 GBd Nyquist QAM-16 and (b) 8 GBd Nyquist QAM-16 signals aggregated from two corresponding Nyquist QPSK signals. Back to back measurement of one of the parent Nyquist QPSK channels resulted in, \textit{Q-factor}$=19.130\pm0.24599$ and \textit{EVM\textsubscript{avg}}$=13.12\pm0.2447$. (c) The measured eye diagram of the in-phase component of the resultant 8 GBd Nyquist QAM-16 signal.}
   \label{fig:sinc2} 
\end{figure}
The symbol constellations and measured performance metrics for the aggregated Nyquist QPSK and PAM-4 data signals from two Nyquist BPSK parent channels with symbol rates 5 and 8 GBd is presented in Fig. \ref{fig:sinc1}. \par
Fig. \ref{fig:sinc2}(a) and \ref{fig:sinc2}(b) presents the symbol constellation along with the corresponding performance metrics for Nyquist QAM-16 signals aggregated from two Nyquist QPSK parent channels with 5 GBd and 8 GBd symbol rates. The eye diagram of the in-phase component of the output 8 GBd Nyquist QAM-16 signal is shown in Fig. \ref{fig:sinc2}(c). The characteristic return to zero nature due to the Nyquist pulse sequence related to the three-line flat phase locked comb is evident from the eye diagram. 

\section{Discussions}
Most of the real-world applications of channel aggregation and modulation format conversion require reconfigurability and scalability in terms of data rates, bandwidth and dynamic range. For the presented technique all three criteria depend on the used modulator and RF source. In contrast to all other aggregation methods presented so far, no optical nonlinearity, special photonic components or pump laser sources are required. Integrated modulators offer broad operational wavelength ranges with very high bandwidth ($\geq100$ GHz) and energy efficiency in many commercially important material platforms like LiNbO3, silicon, and InP \cite{Wang2018,Burla2019,Wang2019,Ogiso2019,Zhang2021}. Therefore, an integration of the method is straightforward on silicon or other material platforms. Although the presented theory and proof-of-concept experiment refer to an MZM, a phase modulator will be sufficient to achieve similar functionality. By cascading the aggregation stages, any number of channels with the aggregated symbol rate can be aggregated to one single or any smaller number of channels with the same symbol rate. \par 
Moreover, phase coherent optical carriers at the transmitter can be sufficed by comb sources like MLLs, microresonators or electro-optic comb generators \cite{Hu2018,Corcoran2020,Kippenberg2011,Zhang2019}. High-capacity coherent optical communication systems have already been demonstrated using such integrated comb sources \cite{Hu2018,Corcoran2020}. Since the exact phase relationship can be adjusted with the electrical frequency applied to the modulator, the only requirement is to have fixed phase relationship between the comb lines at the transmitter.\par
The experimental results shows the feasibility of achieving Nyquist orthogonal time division multiplexing of aggregated higher modulation format channels, to achive higher spectral efficiency \cite{Misra2021,Nakazawa2012, Schmogrow12}. Moreover, more than two channels can also be aggregated by cascading the aggregation stages.

\section{Conclusion}
In conclusion, we have proposed and experimentally demonstrated a linear optical signal processing method to aggregate lower bit rate data channels to fewer higher bit rate channels. The method relies only on electro-optic modulators to manipulate the optical phase by controlling the electrical phase, in order to accomplish a phase coherent vector summation. No optical nonlinearity, pump laser, or special electronics and photonics is needed. Thus, an integration of the method into any material platform is possible. In the first proof-of-concept experiments, we have shown aggregation of 10 GBd NRZ and 8 GBd Nyquist signals of lower modulation formats (e.g. BPSK, QPSK) to spectrally efficient higher order modulation formats (e.g. QPSK, PAM-4, QAM-16) while conserving the symbol rates. 

\section*{Acknowledgments}
This work was supported in part by the Deutsche Forschungsgemeinschaft
(DFG, German Research Foundation) under grant numbers - 322402243,
403154102, 424608109, 424608271, 424607946, 424608191, and in part by
the German Federal Ministry of Education and Research (BMBF) under
funding code 13N14879. We thank Younus Mandalawi and Mohammed Hosni from TU Braunschweig for the constructive discussions. 

\bibliography{report}

\begin{thebibliography}{10}

\bibitem{Amano2019}
Tomoki Amano, Hiroki Kishikawa, and Nobuo Goto.
\newblock {Aggregation of OOK Signals for Modulation Format Conversion to 8QAM
  signal Using XPM and XGM}.
\newblock In {\em OSA Advanced Photonics Congress (AP) 2019 (IPR, Networks,
  NOMA, SPPCom, PVLED)}, page SpM2E.2, Washington, D.C., 2019. OSA.

\bibitem{Burla2019}
Maurizio Burla, Claudia Hoessbacher, Wolfgang Heni, Christian Haffner, Yuriy
  Fedoryshyn, Dominik Werner, Tatsuhiko Watanabe, Hermann Massler, Delwin~L.
  Elder, Larry~R. Dalton, and Juerg Leuthold.
\newblock {500 GHz plasmonic Mach-Zehnder modulator enabling sub-THz microwave
  photonics}.
\newblock {\em APL Photonics}, 4(5):056106, may 2019.

\bibitem{Chitgarha2013}
Mohammad~Reza Chitgarha, Salman Khaleghi, Zahra Bakhtiari, Morteza Ziyadi, Ori
  Gerstel, Loukas Paraschis, Carsten Langrock, Martin~M. Fejer, and Alan~E.
  Willner.
\newblock {Demonstration of reconfigurable optical generation of higher-order
  modulation formats up to 64 QAM using optical nonlinearity}.
\newblock {\em Opt. Lett.}, 38(17):3350--3353, 2013.

\bibitem{Chitgarha2014}
Mohammad~Reza Chitgarha, Salman Khaleghi, Morteza Ziyadi, Ahmed Almaiman,
  Amirhossein Mohajerin-Ariaei, Ori Gerstel, Loukas Paraschis, Carsten
  Langrock, Martin~M. Fejer, Joseph Touch, and Alan~E. Willner.
\newblock {Demonstration of tunable optical generation of higher-order
  modulation formats using nonlinearities and coherent frequency comb}.
\newblock {\em Opt. Lett.}, 39(16):4915--4918, aug 2014.

\bibitem{Corcoran2020}
Bill Corcoran, Mengxi Tan, Xingyuan Xu, Andreas Boes, Jiayang Wu, Thach~G.
  Nguyen, Sai~T. Chu, Brent~E. Little, Roberto Morandotti, Arnan Mitchell, and
  David~J. Moss.
\newblock {Ultra-dense optical data transmission over standard fibre with a
  single chip source}.
\newblock {\em Nat. Commun.}, 11(1):2568, dec 2020.

\bibitem{Fallahpour2020}
Ahmad Fallahpour, Fatemeh Alishahi, Kaiheng Zou, Yinwen Cao, Ahmed Almaiman,
  Arne Kordts, Maxim Karpov, Martin Hubert~Peter Pfeiffer, Karapet Manukyan,
  Huibin Zhou, Peicheng Liao, Cong Liu, Moshe Tur, Tobias~J. Kippenberg, and
  Alan~E. Willner.
\newblock Demonstration of tunable optical aggregation of qpsk to 16-qam over
  optically generated nyquist pulse trains using nonlinear wave mixing and a
  kerr frequency comb.
\newblock {\em Journal of Lightwave Technology}, 38(2):359--365, 2020.

\bibitem{Hu2018}
Hao Hu, Francesco {Da Ros}, Minhao Pu, Feihong Ye, Kasper Ingerslev, Edson
  {Porto da Silva}, Md. Nooruzzaman, Yoshimichi Amma, Yusuke Sasaki, Takayuki
  Mizuno, Yutaka Miyamoto, Luisa Ottaviano, Elizaveta Semenova, Pengyu Guan,
  Darko Zibar, Michael Galili, Kresten Yvind, Toshio Morioka, and Leif~K.
  Oxenl{\o}we.
\newblock {Single-source chip-based frequency comb enabling extreme parallel
  data transmission}.
\newblock {\em Nat. Photonics}, 12(8):469--473, aug 2018.

\bibitem{Huang2011}
Guoxiu Huang, Yuji Miyoshi, Akihiro Maruta, Yuki Yoshida, and Ken-Ichi
  Kitayama.
\newblock {All-Optical OOK to 16-QAM Modulation Format Conversion Employing
  Nonlinear Optical Loop Mirror}.
\newblock {\em J. Light. Technol.}, 30(9):1342--1350, may 2012.

\bibitem{Kippenberg2011}
T.~J. Kippenberg, R.~Holzwarth, and S.~A. Diddams.
\newblock Microresonator-based optical frequency combs.
\newblock {\em Science}, 332(6029):555--559, 2011.

\bibitem{Liu2019a}
Hong Liu, Hongxiang Wang, and Yuefeng Ji.
\newblock {Simultaneous all-optical channel aggregation and De-aggregation for
  8QAM signal in elastic optical networking}.
\newblock {\em IEEE Photonics J.}, 11(1):1--8, 2019.

\bibitem{Liu2019b}
Hong Liu, Hongxiang Wang, Zhen Xing, and Yuefeng Ji.
\newblock {Simultaneous all-optical channel aggregation and de-aggregation
  based on nonlinear effects for OOK and MPSK formats in elastic optical
  networking}.
\newblock {\em Opt. Express}, 27(21):30158--30171, oct 2019.

\bibitem{Lu2015}
Guo-Wei Lu, Jos{\'{e}} Manuel~Delgado Mendinueta, Takahide Sakamoto, Naoya
  Wada, and Tetsuya Kawanishi.
\newblock {Optical 8QAM and 8PSK synthesis by cascading arbitrary 2QAM with
  squared QPSK}.
\newblock {\em Opt. Express}, 23(16):21366--21374, aug 2015.

\bibitem{Misra2021}
Arijit Misra, Janosch Meier, Stefan Preussler, Karanveer Singh, and Thomas
  Schneider.
\newblock {Agnostic sampling transceiver}.
\newblock {\em Opt. Express}, 29(10):14828--14840, may 2021.

\bibitem{Musumeci2016}
Francesco Musumeci, Camilla Bellanzon, Nicola Carapellese, Massimo Tornatore,
  Achille Pattavina, and Stephane Gosselin.
\newblock {Optimal BBU Placement for 5G C-RAN Deployment Over WDM Aggregation
  Networks}.
\newblock {\em J. Light. Technol.}, 34(8):1963--1970, apr 2016.

\bibitem{Nakazawa2012}
Masataka Nakazawa, Toshihiko Hirooka, Peng Ruan, and Pengyu Guan.
\newblock {Ultrahigh-speed “orthogonal” TDM transmission with an optical
  Nyquist pulse train}.
\newblock {\em Opt. Express}, 20(2):1129--1140, jan 2012.

\bibitem{Nakazawa2014}
Masataka Nakazawa, Masato Yoshida, and Toshihiko Hirooka.
\newblock {The Nyquist laser}.
\newblock {\em Optica}, 1(1):15--22, jul 2014.

\bibitem{Ogiso2019}
Y.~Ogiso, J.~Ozaki, Y.~Ueda, H.~Wakita, M.~Nagatani, H.~Yamazaki, M.~Nakamura,
  T.~Kobayashi, S.~Kanazawa, T.~Fujii, Y.~Hashizume, H.~Tanobe, N.~Nunoya,
  M.~Ida, Y.~Miyamoto, and M.~Ishikawa.
\newblock {Ultra-High Bandwidth InP IQ Modulator for Beyond 100-GBd
  transmission}.
\newblock In {\em Opt. Fiber Commun. Conf. 2019}, page M2F.2, Washington, D.C.,
  2019. OSA.

\bibitem{Parmigiani2012}
Francesca Parmigiani, Liam Jones, Joseph Kakande, Periklis Petropoulos, and
  David~J. Richardson.
\newblock {Modulation format conversion employing coherent optical
  superposition}.
\newblock {\em Opt. Express}, 20(26):B322--B330, 2012.

\bibitem{Qiankun2021}
Li~Qiankun, Yang xiong Wei, and Yang Jiali.
\newblock {All-Optical Aggregation and De-aggregation between 8QAM and BPSK
  Signal Based on Nonlinear Effects in HNLF}.
\newblock {\em J. Light. Technol.}, pages 1--1, 2021.

\bibitem{Schmogrow12}
R.~Schmogrow, M.~Winter, M.~Meyer, D.~Hillerkuss, S.~Wolf, B.~Baeuerle,
  A.~Ludwig, B.~Nebendahl, S.~Ben-Ezra, J.~Meyer, M.~Dreschmann, M.~Huebner,
  J.~Becker, C.~Koos, W.~Freude, and J.~Leuthold.
\newblock {Real-time Nyquist pulse generation beyond 100 Gbit/s and its
  relation to OFDM}.
\newblock {\em Optics Express}, 20(1):317--339, jan 2012.

\bibitem{Soto2013}
Marcelo~A. Soto, Mehdi Alem, Mohammad {Amin Shoaie}, Armand Vedadi,
  Camille-Sophie Br{\`{e}}s, Luc Th{\'{e}}venaz, and Thomas Schneider.
\newblock {Optical sinc-shaped Nyquist pulses of exceptional quality}.
\newblock {\em Nat. Commun.}, 4(1):2898, dec 2013.

\bibitem{SotoCleo}
Marcelo~A. Soto, Mehdi Alem, Mohammad~Amin Shoaie, Armand Vedadi,
  Camille-Sophie Br\`{e}s, Luc Th\'{e}venaz, and Thomas Schneider.
\newblock Generation of nyquist sinc pulses using intensity modulators.
\newblock In {\em CLEO: 2013}, page CM4G.3. Optical Society of America, 2013.

\bibitem{Wang2018}
Cheng Wang, Mian Zhang, Xi~Chen, Maxime Bertrand, Amirhassan Shams-Ansari,
  Sethumadhavan Chandrasekhar, Peter Winzer, and Marko Lon{\v{c}}ar.
\newblock {Integrated lithium niobate electro-optic modulators operating at
  CMOS-compatible voltages}.
\newblock {\em Nature}, 562(7725):101--104, 2018.

\bibitem{Wang2019}
Xiaoxi Wang, Peter~O. Weigel, Jie Zhao, Michael Ruesing, and Shayan Mookherjea.
\newblock {Achieving beyond-100-GHz large-signal modulation bandwidth in hybrid
  silicon photonics Mach Zehnder modulators using thin film lithium niobate}.
\newblock {\em APL Photonics}, 4(9):096101, sep 2019.

\bibitem{Willner2019}
Alan~E. Willner, Ahmad Fallahpour, Fatemeh Alishahi, Yinwen Cao, Amirhossein
  Mohajerin-Ariaei, Ahmed Almaiman, Peicheng Liao, Kaiheng Zou, Ari~N. Willner,
  and Moshe Tur.
\newblock {All-Optical Signal Processing Techniques for Flexible Networks}.
\newblock {\em J. Light. Technol.}, 37(1):21--35, jan 2019.

\bibitem{Winzer2012}
Peter~J. Winzer.
\newblock {High-Spectral-Efficiency Optical Modulation Formats}.
\newblock {\em J. Light. Technol.}, 30(24):3824--3835, dec 2012.

\bibitem{Zhang2013}
Banghong Zhang, Hongyu Zhang, Changyuan Yu, Xiaofei Cheng, Yong~Kee Yeo,
  Pooi-Kuen Kam, Jing Yang, Hongguang Zhang, Yu-Hsiang Wen, and Kai-Ming Feng.
\newblock {An All-Optical Modulation Format Conversion for 8QAM Based on FWM in
  HNLF}.
\newblock {\em IEEE Photonics Technol. Lett.}, 25(4):327--330, feb 2013.

\bibitem{Zhang2019}
Mian Zhang, Brandon Buscaino, Cheng Wang, Amirhassan Shams-Ansari, Christian
  Reimer, Rongrong Zhu, Joseph~M. Kahn, and Marko Lon{\v{c}}ar.
\newblock {Broadband electro-optic frequency comb generation in a lithium
  niobate microring resonator}.
\newblock {\em Nature}, 568(7752):373--377, apr 2019.

\bibitem{Zhang2021}
Mian Zhang, Cheng Wang, Prashanta Kharel, Di~Zhu, and Marko Lon{\v{c}}ar.
\newblock {Integrated lithium niobate electro-optic modulators: when
  performance meets scalability}.
\newblock {\em Optica}, 8(5):652, may 2021.

\end{thebibliography}
\bibliographystyle{plain}

\end{document}